\documentclass[reprint, amsmath,amssymb, pra]{revtex4-2}
 \usepackage[utf8]{inputenc}
\usepackage{xcolor}
\usepackage{xcolor}
\newcommand{\N}{\text{N}}
\newcommand{\No}{\text{N}}
\usepackage[
  colorlinks=true,
  linkcolor=red,
  citecolor=red,
  urlcolor=red 
]{hyperref}
\usepackage{mathrsfs}
\usepackage{graphicx}
\usepackage{bm}
\usepackage{orcidlink}
\usepackage{appendix}
\usepackage{braket}
\begin{document}

    \title{Higher-order nonclassicality criteria for photon-subtracted and photon-added states via the normalization constant}

\author{Jhordan Santiago\,\orcidlink{0009-0001-9265-9827}}
 \email{student.jhordansantiago@gmail.com}
\affiliation{Unidade Acadêmica de Física, Universidade Federal de Campina Grande, 10071, 58429-900, Campina Grande, Brazil.
 }

\date{\today}
\begin{abstract}
We show that any nonclassicality criterion based on factorial moments, including several higher-order parameters such as the Mandel $Q^{(\ell)}$ parameter, the Lee antibunching function $d^{(\ell-1)}_h$, and the Agarwal--Tara parameter $A_3$, can be computed straightforwardly for photon-subtracted and photon-added states by performing operator reordering of the factorial moments. Within this approach, all relevant quantities depend solely on the normalization constant of the given state.
\end{abstract}

\maketitle
\section{Introduction}

The rigorous quantification of higher-order nonclassicality typically involves the calculation of high-order factorial moments of the photon-number distribution, which can become algebraically complex for non-trivial states. Criteria such as the generalized Mandel $Q^{(\ell)}$ parameter \cite{Kim2002}, the Lee antibunching function $d^{(\ell-1)}_h$ \cite{Lee1990}, and the Agarwal-Tara parameter $A_3$ \cite{Agarwal1992}, are some examples of tools for such analyzes. A simplified yet comprehensive method for evaluating these criteria is highly desirable to facilitate the characterization and study of such states. 

In this Letter, we present a  straightforward approach to compute these key higher-order nonclassicality criteria for any photon-subtracted or photon-added state. By expressing the factorial moments in terms of the anti-normal and normal ordering of the field operators, we demonstrate that all necessary moments can be derived solely from the state's normalization constant. 

\section{Nonclassicality}\label{non}
The criteria discussed in this work are all based on the factorial moments of the photon-number distribution $\langle a^\dagger a \rangle$, where $a^\dagger$ ($a$) is the creation (annihilation) operator. In what follows, we briefly discuss some of them.

\subsection{Generalized Mandel $Q^{(\ell)}$ parameter}

The Mandel $Q$ parameter \cite{Fano1947, Mandel1979} quantifies the deviation of the photon-number variance from the mean. It is defined as
\begin{align}
Q
= \frac{\langle (a^\dagger a)^2 \rangle 
      - \langle a^\dagger a \rangle^2
      - \langle a^\dagger a \rangle}
     {\langle a^\dagger a \rangle},
\end{align}
where $Q < 0$ indicates sub-Poissonian statistics.

The generalized Mandel parameter $Q^{(\ell)}$ \cite{Kim2002} extends this concept to higher orders.  
Introducing  
\begin{align}
\Delta n = a^\dagger a - \langle a^\dagger a \rangle,
\end{align}
the parameter becomes
\begin{align}
Q^{(\ell)}
&\equiv 
\frac{\left\langle (\Delta n)^{2\ell} \right\rangle
      - O_{2\ell}\!\left(\langle a^\dagger a \rangle\right)}
     {O_{2\ell}\!\left(\langle a^\dagger a \rangle\right)}
\\[4pt]
&=
\frac{\left\langle :(\Delta n)^{2\ell}: \right\rangle}
     {O_{2\ell}\!\left(\langle a^\dagger a \rangle\right)},
\qquad \ell = 1,2,3,\dots
\end{align}
where $O_{2\ell}(\langle a^\dagger a \rangle)$ is the value that  
$\langle(\Delta n)^{2\ell}\rangle$ assumes for a Poisson distribution with mean  
$\langle a^\dagger a \rangle$.

\subsection{Lee antibunching function $d^{(\ell-1)}_h$}
Although referred to as an antibunching criterion, the function $d^{(\ell-1)}_h$ introduced by Lee \cite{Lee1990}, and later refined to gain a clearer physical meaning in \cite{Pathak2006}, is fundamentally a measure of higher-order sub-Poisson statistics, as it depends only on factorial moments and not on time-dependent correlation functions. A nonclassical state can exhibit any combination of (anti)bunching and sub(super)-Poisson statistics, since one does not imply the other \cite{Singh1983,Zou1990,Mandel1995,Emary2012}.
 The function $d^{(\ell-1)}_h$ generalizes the concept of sub-Poisson statistics to higher orders $\ell$. It is defined as the difference between the $\ell$-th order factorial moment and the $\ell$-th power of the first-order moment:
\begin{equation}
d^{(\ell-1)}_h = \langle a^{\dagger \ell} a^\ell \rangle - \langle a^\dagger a \rangle^\ell.
\end{equation}
The condition for higher-order sub-Poissonian statistics, and thus nonclassicality, is $d^{(\ell-1)}_h < 0$.

\subsection{Agarwal-Tara parameter $A_3$}
The Agarwal-Tara parameter $A_3$ is a specific higher-order nonclassicality criterion that involves up to fourth-order factorial moments \cite{Agarwal1992}. It was introduced to detect nonclassicality in states that do not exhibit sub-Poisson statistics nor quadrature squeezing. The parameter is defined as:
\begin{align}
A_{3}=\frac{\operatorname{det} m^{(3)}}{\operatorname{det} \mu^{(3)}-\operatorname{det} m^{(3)}}<0
\end{align}

where \begin{align*}m^{(3)}=\left[\begin{array}{lll}1 & m_{1} & m_{2} \\ m_{1} & m_{2} & m_{3} \\ m_{2} & m_{3} & m_{4}\end{array}\right] \quad \text { and } \quad \mu^{(3)}=\left[\begin{array}{lll}1 & \mu_{1} & \mu_{2} \\ \mu_{1} & \mu_{2} & \mu_{3} \\ \mu_{2} & \mu_{3} & \mu_{4}\end{array}\right].
\end{align*}
with  $m_{x}=\left\langle a^{\dagger x} a^{x}\right\rangle$ and $\mu_{z}=\left\langle\left(a^{\dagger} a\right)^{z}\right\rangle$ are the matrix elements.

\section{Photon-subtracted states}
Let’s consider a normalized photon-subtracted state defined as
\begin{equation}
|\psi^{(n)}\rangle = \No_n^{-1/2} a^n |\psi\rangle,
\label{eq:subtracted_state}
\end{equation}
where $n$ is the number of photon subtractions and the normalization constant $\No_n$ ensures $\langle \psi^{(n)} | \psi^{(n)} \rangle = 1$, yielding

\begin{equation}
    \No_n = \langle \psi | a^{\dagger n} a^n | \psi \rangle.
\end{equation}

The expectation value $\langle a^{\dagger x} a^x \rangle$ is easily computed because it is already in the normally order:
\begin{align}
\langle a^{\dagger x} a^x \rangle &= \langle \psi^{(n)} | a^{\dagger x} a^x | \psi^{(n)} \rangle \label{eq:axax_subtracted_1} \\
&= \No_n^{-1} \langle \psi | a^{\dagger n} a^{\dagger x} a^x a^n | \psi \rangle\nonumber\\
&= \No_n^{-1} \langle \psi | a^{\dagger (n+x)} a^{n+x} | \psi \rangle\nonumber\\& = \frac{\No_{n+x}}{\No_n}.\label{eq:sub0}
\end{align}

From Eq. \eqref{eq:sub0}, any criterion based on factorial moments can be directly calculated. For example, the  Mandel $Q$ parameter reduces to:
\begin{align}
    Q=\frac{\N_{m+2}}{\N_{m+1}}-\frac{\N_{m+1}}{\N_m}, 
\end{align}
and Lee's antibuching function yields
\begin{align}
    d_h^{(x-1)}=\frac{\N_{n+x}}{\N_n}-\left(\frac{\N_{n+1}}{\N_n}\right)^x.
\end{align}

\section{Photon-added states}

We shall now consider a normalized photon-added state
\begin{equation}
|\psi^{(m)}\rangle = \N_m^{-1/2} a^{\dagger m} |\psi\rangle,
\label{eq:added_state}
\end{equation}
whose normalization constant is
\begin{equation}
\N_m = \langle \psi | a^m a^{\dagger m} | \psi \rangle .
\label{eq:N_m}
\end{equation}
Using the operator identity
\begin{equation}
a^{\dagger x} a^{x}
= \sum_{k=0}^{x}
(-1)^k k! \binom{x}{k}^2
a^{x-k} a^{\dagger(x-k)},
\label{eq:identity}
\end{equation}
and a similar property as of Eq. \eqref{eq:sub0}, but for the photon-added state, we have
\begin{align}
    \langle a^{x-k}a^{\dagger (x-k)}\rangle&=\langle\psi|a^m  a^{x-k}a^{\dagger (x-k)}a^{\dagger m}|\psi\rangle\nonumber\\
    &=\frac{\N_{m+x-k}}{\N_m},\label{s}
\end{align}
with $m\geq x-k$ (in a broader sense, Eq. \eqref{s} is not strictly constrained by the condition $m \geq x-k$, since inverse operators of the type $a^{\dagger -y}$, where $y$ is a number, can be constructed \cite{Mehta1992}. Nevertheless, in the present work we restrict our analysis to the conventional framework commonly employed in the literature), hence

\begin{equation}
\langle a^{\dagger x} a^x \rangle
= \frac{1}{\N_m}
\sum_{k=0}^x (-1)^k k! \binom{x}{k}^2 \N_{m+x-k},
\label{eq:axax_added_final_sum}
\end{equation}
which depends only on the normalization constant of the photon-added state.

As an example, the Mandel $Q$ parameter for the photon-added state becomes
\begin{equation}
Q
= \frac{\N_{m+2}-4\N_{m+1}+2\N_m}{\N_{m+1}-\N_m}
  - \frac{\N_{m+1}-\N_m}{\N_m}.
\label{eq:mandel_added}
\end{equation}

Eq. \eqref{eq:mandel_added} is not new; it was previously obtained, for example, in Ref. \cite{Ren2014} using the integration within an ordered product (IWOP) technique.

\section{Discussion}
In this work, we presented a straightforward approach to evaluate several higher-order nonclassicality criteria for both photon-subtracted and photon-added states. The core of our method lies in showing that all required factorial moments can be obtained directly from the state's normalization constant. This follows from expressing the expectation values in anti-normal and normal order. In the case of photon-subtracted states, which are already in normally ordered form by construction, the resulting expressions become particularly simple and compact.

\end{document}